\documentclass[11pt,twoside]{article}
\usepackage{asp2014}

\aspSuppressVolSlug
\resetcounters

\begin{document}

\title{OPUS: an interoperable job control system based on VO standards}

\author{Mathieu~Servillat$^1$ and Stéphane~Aicardi$^2$ and Baptiste~Cecconi$^3$ and Marco~Mancini$^4$}
\affil{$^1$LUTH, Observatoire de Paris, CNRS, PSL, 92190 Meudon, France; \email{mathieu.servillat@obspm.fr}}
\affil{$^2$DIO, Observatoire de Paris, PSL, 92190 Meudon, France}
\affil{$^2$LESIA, Observatoire de Paris, CNRS, PSL, 92190 Meudon, France}
\affil{$^2$Institut Denis Poisson, CNRS, 37200 Tours, France}

\paperauthor{Mathieu~Servillat}{mathieu.servillat@obspm.fr}{0000-0001-5443-4128}{LUTH, Observatoire de Paris, CNRS, PSL}{}{Meudon}{}{92190}{France}
\paperauthor{Stéphane~Aicard}{}{0000-0003-4475-0930}{DIO, Observatoire de Paris, CNRS, PSL}{}{Meudon}{}{92190}{France}
\paperauthor{Baptiste~Cecconi}{}{0000-0001-7915-5571}{LESIA, Observatoire de Paris, CNRS, PSL}{}{Meudon}{}{92190}{France}
\paperauthor{Marco~Mancini}{}{}{Institut Denis Poisson, CNRS}{}{Tours}{}{37200}{France}



  
\begin{abstract}

OPUS (Observatoire de Paris UWS System) is a job control system that aims at facilitating the access to analysis and simulation codes through an interoperable interface. The Universal Worker System pattern v1.1 (UWS) as defined by the International Virtual Observatory Alliance (IVOA) is implemented as a REST service to control the asynchronous execution of a job on a work cluster. OPUS also follows the recent IVOA Provenance Data Model recommendation to capture and expose the provenance information of jobs and results. By following well defined standards, the tool is interoperable and jobs can be run either through a web interface, or a script, and can be integrated to existing web platforms. Current instances are used in production by several projects at the Observatoire de Paris (CTA/H.E.S.S, MASER, CompOSE).
  
\end{abstract}

\section{Objective and use cases}

Research teams at the Observatoire de Paris often develop customized analysis and simulation codes for their projects. They also have access to locally managed computing resources. However, the visibility of the codes sometimes remains low, with a sub-optimal use of the available resources. The objective of OPUS is to provide a simple and interoperable access to those codes, from either a small team or a larger public, on either a local machine or a work cluster. OPUS executes jobs asynchronously to enable the management of jobs with long execution duration.

OPUS is developed in the context of the CTA and H.E.S.S projects at the LUTH\footnote{\url{https://voparis-cta-confluence.obspm.fr/display/CD/Analysis+tools}}. It is tested with public H.E.S.S. data processed and analysed with  tools such as \texttt{gammapy}, \texttt{ctools} or \texttt{ctapipe} that run on the local cluster. The MASER project provides a tool\footnote{\url{https://maser.lesia.obspm.fr/task-2-modeling-tools/expres}} that allows to simulate auroral radio emission dynamic spectra \citep{2019A&A...627A..30L}. The code is available for run-on-demand operations through an OPUS instance. The online service CompOSE\footnote{\url{https://compose.obspm.fr}} \citep{2013arXiv1307.5715T} provides data tables for different state of the art equations of state for dense matter. Those data tables can be generated on demand through an OPUS instance.

\section{Modularity}

The application is composed of several modules  connected by different Application Programming Interfaces (see Figure~\ref{fig1}) : i/ a \textbf{Web Server} that receives job control commands, stores job information and manages jobs on ii/ the \textbf{Work Cluster} that executes the jobs; and iii/ a \textbf{Web Client} that shows job lists, job details with result preview (see snapshots in Figure~\ref{fig2}), and which includes a job definition editor and an admin interface.

Those modules may be executed on different machines or on the same laptop. The documentation\footnote{\url{https://opus-job-manager.readthedocs.io}} indicates how to install and configure the system. The source code is available on GitHub\footnote{\url{https://github.com/ParisAstronomicalDataCentre/OPUS}}.

\articlefigure{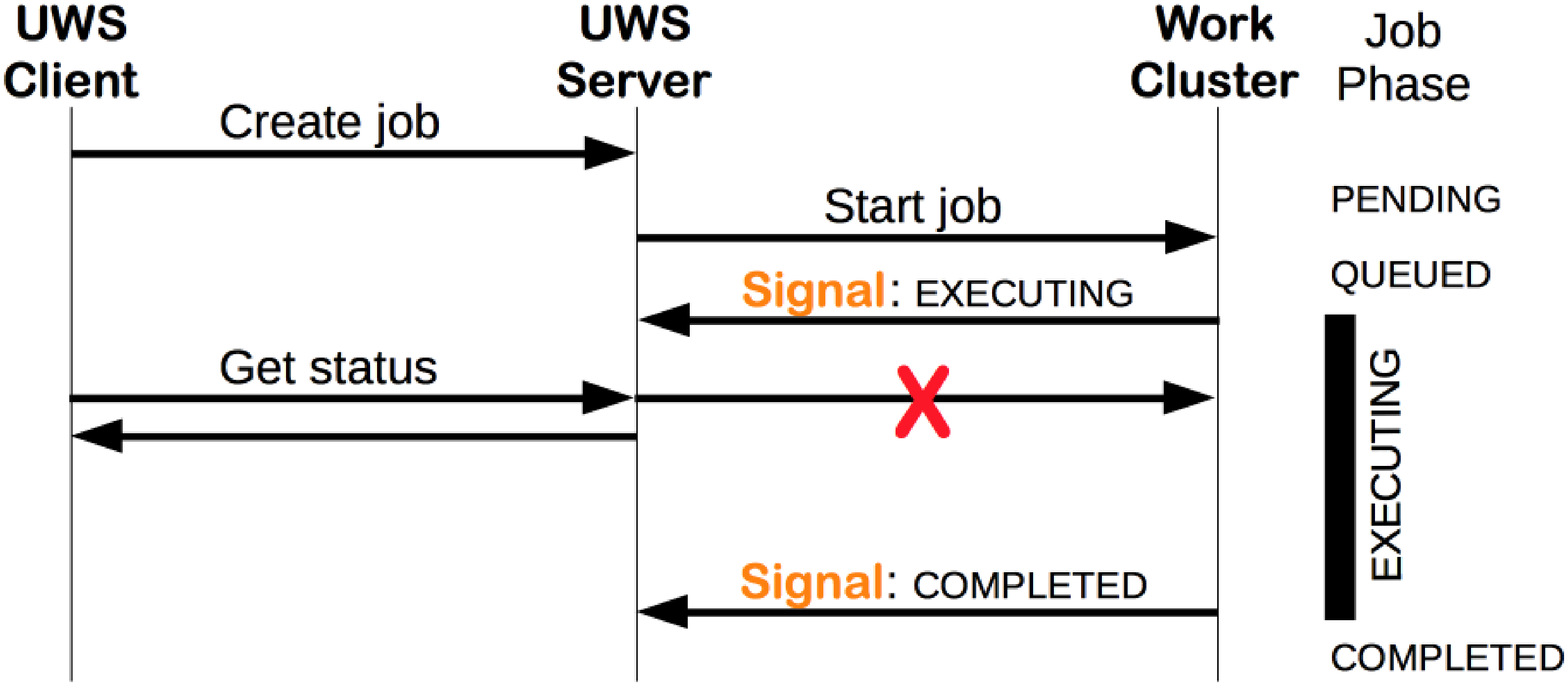}{fig1}{Communications between OPUS modules.}

\articlefigure{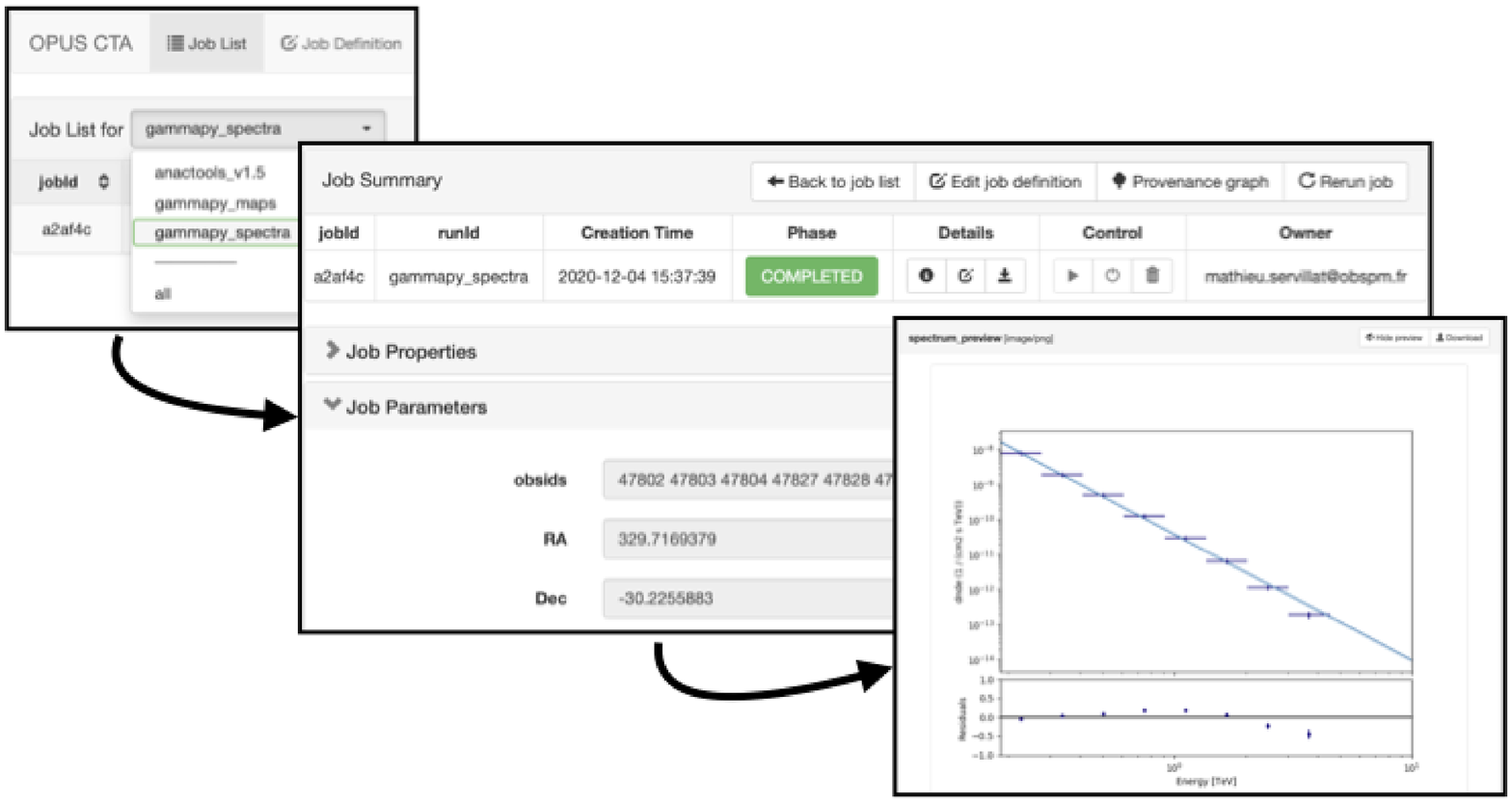}{fig2}{Web client pages to select a job, view it and visualize a result.}

\section{Interfaces and standards}

The IVOA Universal Worker System pattern\footnote{\url{https://www.ivoa.net/documents/UWS}} (UWS, \citealt{2016ivoa.spec.1024H}) defines how to manage asynchronous execution of jobs on a service. The Web Server implements UWS as a REST API and returns standard XML outputs. The UWS commands may be sent by the Web Client or using scripts and command lines.

The IVOA Provenance Data Model\footnote{\url{https://www.ivoa.net/documents/ProvenanceDM}} (ProvDM, \citealt{2020ivoa.spec.0411S}) is used for job definitions and provenance tracing. The job definition editor implements the ProvDM description classes (Usage, Generation, Parameters…). A ProvSAP (Simple Access Protocol) interface returns graphs built with the \texttt{voprov} Python package. An example of the output is presented in Figure~\ref{fig3}.

In order to manipulate the job definitions, the IVOA service descriptor serialization is extended to include descriptions of usage and generation of entities by the job. Such files are VOTable composed of a RESOURCE block (\texttt{type}~= meta, \texttt{utype}~= voprov:ActivityDescription), that contains attributes for the activity description as PARAM blocks, and then GROUP blocks for InputParams, Usage and Generation. This transposes to a JSON/YAML file exchanged between web services, a job activity being described by a dictionary of attributes, that includes sub-dictionaries for parameters, usage and generation.

The Server interfaces with a Work Cluster and its batch queue system (SLURM, or local jobs are currently supported). User accounts on the Server are managed from the Client using the System for Cross-domain Identity Management (SCIM) standards\footnote{\url{http://www.simplecloud.info}}.

\section{Next developments}

OPUS is also a test-bed to develop new functionalities related to the management of provenance information. In particular, the granularity of the provenance can be refined by providing detailed internal provenance at the end of the job, e.g. generated with the \texttt{logprov} Python package\footnote{\url{https://github.com/mservillat/logprov}} that uses a similar YAML serialization for job or activity definitions.

In order to better integrate OPUS with other services, we plan to include federation authentication support (e.g. SAML or OpenID Connect) to the authentication system.

It is also foreseen to provide a Docker container to deploy the system more easily, with an automatic setup of the environment.

\articlefigure{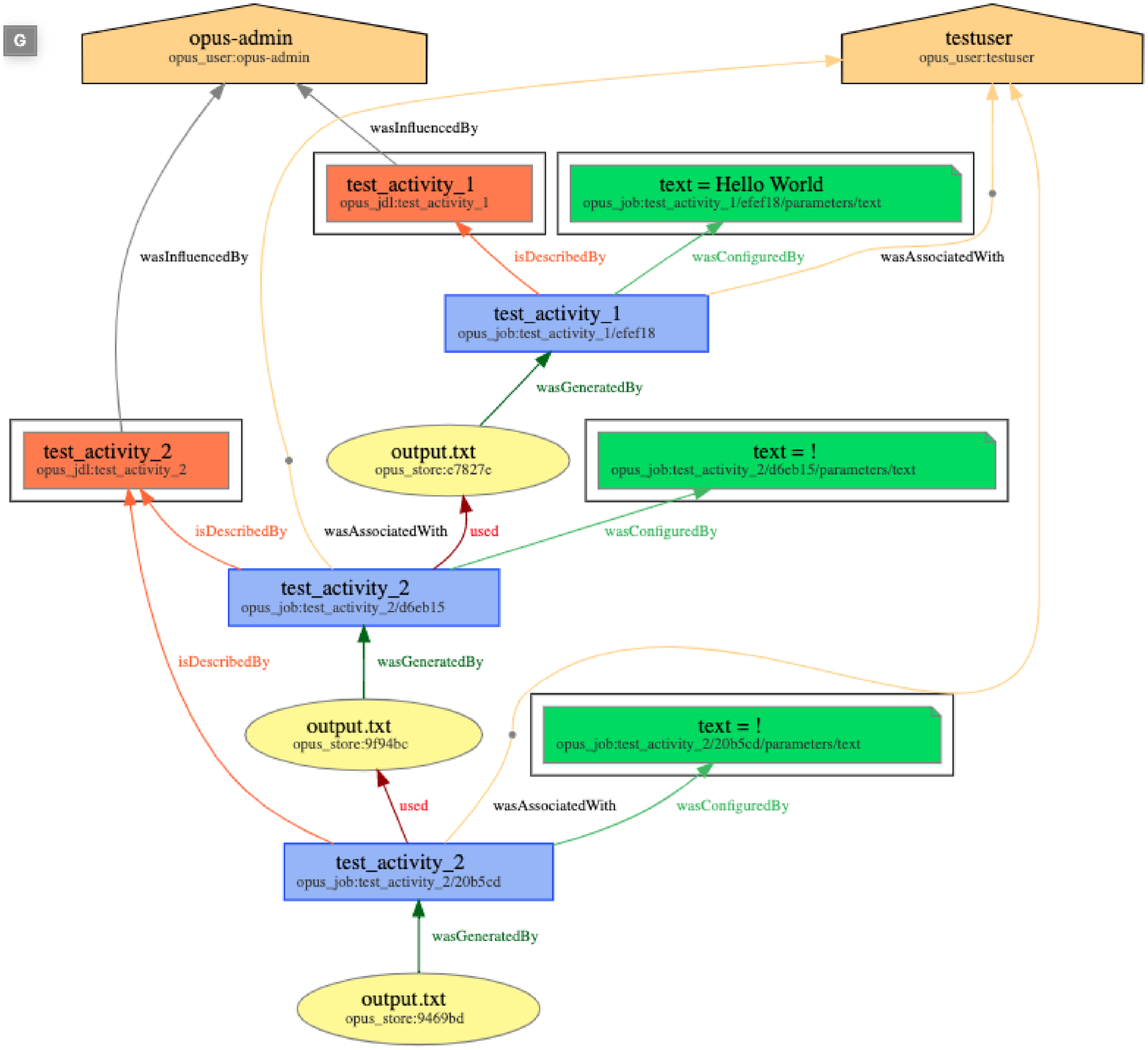}{fig3}{Provenance graph returned by OPUS for a sequence of test jobs. Jobs (Activities) are in blue square boxes, files (Entities) in yellow round shapes and users (Agents) in light orange boxes. Description classes are shown in dark orange, and configuration parameters in green.}


\bibliography{P9-89}


\end{document}